\documentclass[reprint,superscriptaddress,amsmath,amssymb,aps,prb]{revtex4-2}
\pdfoutput=1
\usepackage{graphicx}
\usepackage{dcolumn}
\usepackage{bm}
\usepackage{hyperref}
\usepackage{units}
\usepackage[english]{babel}
\usepackage{color}

\usepackage{mathtools}
\usepackage{dsfont}

\begin{document}

\title{Citation Sentiment Reflects Multiscale Sociocultural Norms}

\author{Xiaohuan Xia}
\affiliation{
Department of Bioengineering, University of Pennsylvania, Philadelphia, PA 19104 USA
}
\author{Mathieu Ouellet}
\affiliation{
Department of Electrical \& Systems Engineering, University of Pennsylvania, Philadelphia, PA 19104 USA
}
\author{Shubhankar P. Patankar}
\affiliation{
Department of Bioengineering, University of Pennsylvania, Philadelphia, PA 19104 USA
}

\author{Diana I. Tamir}
\affiliation{
Department of Psychology, Princeton University, Princeton, NJ 08544 USA
}

\author{Dani S. Bassett}
\email{Corresponding author: dsb@seas.upenn.edu}
\affiliation{
Department of Bioengineering, University of Pennsylvania, Philadelphia, PA 19104 USA
}
\affiliation{
Department of Physics and Astronomy, University of Pennsylvania, Philadelphia, PA 19104 USA
}
\affiliation{
Department of Electrical \& Systems Engineering, University of Pennsylvania, Philadelphia, PA 19104 USA
}
\affiliation{
Department of Neurology, University of Pennsylvania, Philadelphia, PA 19104 USA
}
\affiliation{
Department of Psychiatry, University of Pennsylvania, Philadelphia, PA 19104 USA
}
\affiliation{
Santa Fe Institute, Santa Fe, NM 87501 USA
}
\affiliation{
Montreal Neurological Institute, McGill University, Montreal, QC H3A 2B4, Canada
}

\date{\today}

\begin{abstract}
Modern science is formally structured around scholarly publication, where scientific knowledge is canonized through citation. Precisely how citations are given and accrued can provide information about the value of discovery, the history of scientific ideas, the structure of fields, and the space or scope of inquiry. Yet parsing this information has been challenging because citations are not simply present or absent; rather, they differ in purpose, function, and sentiment. In this paper, we investigate how critical and favorable sentiments are distributed across citations, and demonstrate that citation sentiment tracks sociocultural norms across scales of collaboration, discipline, and country. At the smallest scale of individuals, we find that researchers cite scholars they have collaborated with more favorably (and less critically) than scholars they have not collaborated with. Outside collaborative relationships, higher h-index scholars cite lower h-index scholars more critically. At the mesoscale of disciplines, we find that wetlab disciplines tend to be less critical than drylab disciplines, and disciplines that engage in more synthesis through publishing more review articles tend to be less critical. At the largest scale of countries, we find that greater individualism (and lesser acceptance of the unequal distribution of power) is associated with more critical sentiment. Collectively, our results demonstrate how sociocultural factors can explain variations in sentiment in scientific communication. As such, our study contributes to a broader understanding of how human factors influence the practice of science, and underscores the importance of considering the larger sociocultural contexts in which science progresses.

\end{abstract}

\keywords{Sentiment Analysis $|$ Ingroup Relations $|$ Citation Practices $|$ Coauthorship}
\maketitle

\section{Introduction}

In the modern era, science has often been assumed to be an objective process \cite{datson2010objectivity}. By this descriptor, scholars have variously meant faithfulness to facts, absence of normative commitments, and freedom from personal biases \cite{reiss2020scientific}. Yet with an increasing understanding of human psychology has come an acknowledgment that even the most scrupulous of researchers cannot achieve various types of objectivity; for example, (1) the mechanical objectivity of suppressing the universal human propensity to judge and aestheticize, or (2) the aperspectival objectivity of eliminating individual idiosyncracies \cite{daston1992objectivity}. The growing appreciation of human subjectivity has motivated a turn from the personal to the community, where objectivity is considered a quality that characterizes a collection or population of studies \cite{freese2018emergence}, and hence is a feature of scientific communities and their practices \cite{douglas2011facts}. In communities, objectivity can further be thought of as occurring in degrees, such that any method of inquiry is only objective to the degree that it permits \emph{transformative criticism} through shared standards, equality of intellectual authority, avenues for criticism, and uptake of criticism \cite{longino1990science}.

% Fig. 1 ================================================================================================
\begin{figure}[t]
    \centering
	\includegraphics[width=\columnwidth]{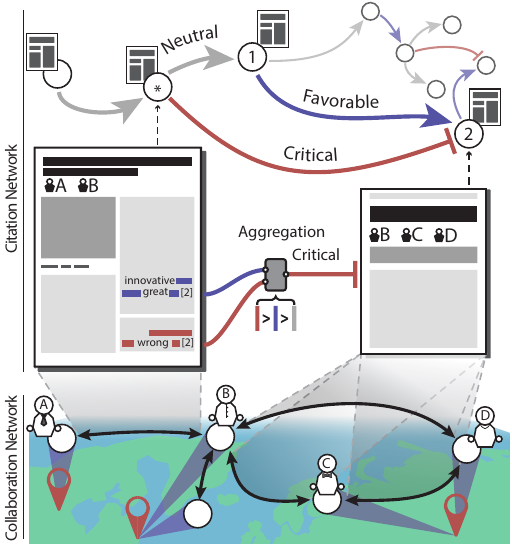}
	\caption{\textbf{Sentiment analysis in citation and author networks.} 
    A two-layer network comprising a network of citations and a network of collaborators. In the citation network, nodes represent scientific articles, and edges represent citations. Sentiment analysis categorizes each edge as favorable (blue), neutral (grey), or critical (red).
    In the collaboration network, nodes represent scholars, and scholars who co-author articles are connected by edges.
	\label{Fig1}}
\end{figure}

Yet, the above discourse begs the question: Need objectivity always be fundamentally at odds with subjectivity? In a complementary line of work, it has been noted that strong objectivity can be maintained without the requirement of value-neutrality \cite{harding2015objectivity} or the suppression of all aspects of the self \cite{sheppard2022subjectivities}, including one's epistemological standpoint \cite{harding1992rethinking} or ethical commitments \cite{resnik2016ethical}. The potential coexistence of certain objectivities and certain subjectivities has proven useful in evaluating the scientific self, which exists among values, standpoints, cognitive biases, and human perceptions \cite{harding2015after}. Indeed, scientists specifically, and scholars more broadly, exist in multiply-dependent cultural contexts---from the local collaborative group or research institution to the scientific discipline or the geographically-defined culture---that each evince distinct norms, values, currencies, and artifacts \cite{crandall2003scientists}. For example, separate disciplines recommend distinct forms of scientific rationality, methods to investigate the world, and factors that others in the community would appreciate \cite{daston2016cloud}; separate countries can have social and funding norms that impact collaborative tendencies \cite{gazni2012mapping}; and even different fields of thought work under distinct notions of theoretical aesthetics, the value of parsimony, and preferred characteristics of explanatory structures \cite{gilbert2021subjectifying}.

Hence, scientific selves are not fundamentally autonomous, self-creating, culture-free individuals; rather, they are co-produced by interactions within the networks, communities, and social movements of scientific culture \cite{harding2015after}, with its set of standards, mores, practices, languages, and dialects \cite{crandall2003scientists}. A key locus of this cultural production is the accumulation of cultural archives through the publication of scientific papers \cite{crandall2003scientists}. Although a large fraction of such papers will not be referenced (or quickly become obsolete \cite{lariviere2008long}), the remainder will be canonized into scientific knowledge proper through the act of citation \cite{lariviere2009decline}. Factors that determine which papers get cited include epistemic values (e.g., empirical support, simplicity, generality, precision, rigor, testability, and explanatory power \cite{longino1990science,budd1999citations}) and non-epistemic values (e.g., common interest, shared gender, same ethnicity, social capital, and institutional privilege \cite{bruggeman2012detecting,mitra2020development,ray2024citation,greenwald1994ethnic,li2013coauthorship,morgan2018prestige}), as well as seemingly trivial factors such as punctuated titles \cite{whissell1999linguistic}. Collectively, these factors provide an explanation for who cites who, how often, and when. In other words, citation dynamics are socio-cognitive processes \cite{espinosa2024coevolution}. 

Citation networks have played a key role in the sociology of science and its critical evaluation \cite{teich2022citation,morgan2018prestige,hicks1991sociology,restivo1987critical}. Yet, due to various challenges of data accessibility and limitations in algorithmic efficacy, a major gap in knowledge exists: precisely how, or in what way, do we cite? Citations can be differently valued \cite{giuffrida2019are}; can be epistemic or procedural \cite{budd1999citations}; can be positive or negative, indicate either support or contrast, attribute creativity or indicate status as a ``classic'' \cite{shadish1995author,safer2009psychology}; can either represent or misrepresent the work cited \cite{roth2010referencing}; and can be chosen for social reasons (such as name-dropping \cite{roth2010referencing} or soliciting favor from a scholar who might have influence in the review process), for mercantile reasons (such as bartering for a citation in return), and for alignment reasons (such as transmitting a specific self-image, e.g., as part of the mainstream, the \emph{avant garde}, or a given school) \cite{erikson2014taxonomy,saez2019quanlitative}. Because not all citations are created equal, citation analysis could be meaningfully expanded to account for citation type \cite{hicks1991sociology,erikson2014taxonomy} and context \cite{small2004shoulders}. Although a few early studies attempted such an expansion, most methods are limited in performance \cite{yousif2019survey} and few use deep learning \cite{nicholson2021scite,ravi2018article} or large language models \cite{hariri2024sentiment} to assess citation sentiment in a more comprehensive and efficient manner.

Here, we overcome this fundamental limitation by leveraging large language models to classify citation sentiment while examining its dependence upon the sociocultural factors in which science exists and progresses (Fig.~\ref{Fig1}). To ensure that our assessment is computationally tractable while remaining sensitive to local cultural norms, we focus on a single discipline---neuroscience---and build a large database of papers from $56,034$ last authors published in $181$ journals (impact factor $\geq 3$ in 2022). As a candidate field, neuroscience has particular affordances for our purpose: its relative youth allows for the emergence and growth of not-yet-crystallized subdisciplines, a fundamental interdisciplinarity as it draws on earlier-defined disciplines, and a marked intersection of opinions, disciplinary methods, and approaches to study \cite{sabbatini2002interdisciplinarity,bilyk2022neurobiology}. All of these factors can support a diversity, heterogeneity, and complexity of the space of ideas, and the network of citations among papers espousing those ideas. We chose to focus on citation sentiment (neutral, favorable, or critical), as a kind or type, for functional and pragmatic purposes. Functionally, sentiment can track coalescence (or fragmentation) of people and ideas; pragmatically, sentiment can be measured quantitatively in extensive databases using large language models. Using this approach, we show how citation sentiment varies across people groups, scientific disciplines, and whole countries, providing a paper-trail lens into the socio-cognitive processes of science.

% Fig. 2 ================================================================================================
\begin{figure}[t]
    \centering
	\includegraphics[width=\columnwidth]{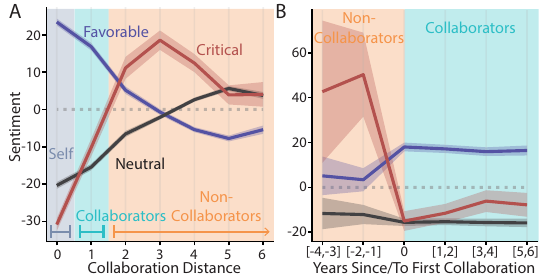}
	\caption{\textbf{Citation sentiment depends on distance and years since (or to) first collaboration.}
        \emph{(A)} Citation sentiment relative to the null model as a function of the collaboration distance between the last authors of the citing and cited works. Shaded regions denote one standard deviation from the mean at a given collaboration distance.
        \emph{(B)} Citation sentiment as a function of the number of years since the first collaboration between the two last authors. A negative year implies that at the time of a citation, the two scholars had yet to collaborate but would do so in the future. Shaded regions denote one standard deviation.
        }
	\label{Fig2}
\end{figure}

\section{Results}

In a large connected citation network of $108909$ papers from the PMC Open Access Subset, we assess citation sentiment as neutral, favorable, or critical (see Supplement). Neutral sentiment is observed in factual phrases or statements, devoid of opinion (e.g., “was reported”, “has been shown”, “can alter”, “can play a role”, “is”, “responds to”). Favorable sentiment is observed in phrases or statements that indicate a positive assessment or evaluation (e.g., adjectival phrases like "elegantly shown", “optimized method”, “powerful tools”) or that indicate consensus or consistency (e.g., “converging results”, “conclusion is consistent”). Critical sentiment is observed in phrases or statements that indicate debate (“the focus of debate”), disagreement (“while other studies do X, we argue Y”), difference (“in contrast to a previous report”), distinction (“is not applicable here”), distance (“despite the fact that many studies have evaluated X, very few have evaluated Y”), or limitation (“however, X method limits desirable outcome Y”). In this formulation, then, sentiment captures both social and epistemic alignment by measuring evaluation and intellectual commonality, respectively.

After measuring the sentiment of $627108$ extracted citation sentences, counted by the number of citation locations in a sentence, we assign sentiment values to each of the citer-citee pairs and aggregate them (see Methods and Supplement for further details). We observe that $42.21\%$ are favorable, $7.94\%$ are critical, and $49.85\%$ are neutral. Our first question is how these sentiments align with ingroup/outgroup relations and structures of dominance. Specifically, we investigate how sentiment is assigned to the ingroup of collaborators (vs. the outgroup of scholars one has never co-authored with) and to the dominant group of scholars with high h-indices (vs. the less dominant group of scholars with lower h-indices).

% Fig. 3 ================================================================================================
\begin{figure}[t]
    \centering
	\includegraphics[width=\columnwidth]{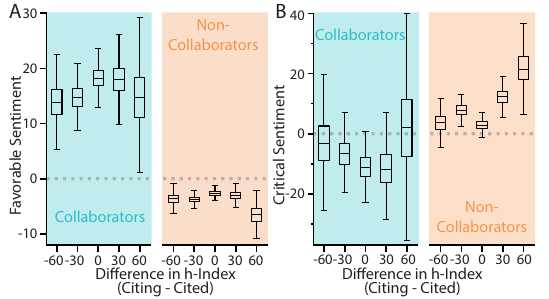}
	\caption{\textbf{Higher h-index scholars critically cite lower h-index scholars.} Citation sentiment as a function of the difference in h-index between the citing and cited last authors.
        }
	\label{Fig3}
\end{figure}

\subsection{Citation Sentiment is Most Favorable to Collaborators}

To examine how sentiment might reflect ingroup/outgroup relations, we measured the collaboration distance: the geodesic distance in the collaboration network between the citer and the citee \cite{newman2001best} at the year of citation. We observed that people are most favorable to themselves and collaborators, and most critical to scholars with whom they have not collaborated (Fig.~\ref{Fig2}A). In subsequent analyses, we removed self-citations to focus on community practices. Note that collaboration distance likely tracks both social distance and intellectual distance. What happens when intellectual distance is held fixed, and social distance changes? To take an initial step toward addressing this question, we isolate citer-citee pairs who would eventually become collaborators. Just prior to collaborating, we find that sentiment is overwhelmingly critical, whereas just after collaborating, critical sentiment plummets and favorable sentiment rises (Fig.~\ref{Fig2}B). This pattern suggests that the social act of collaboration may drive a transformation from critical to favorable sentiment among scholars. We also notice that citer-citee pairs that will collaborate within the next four years tend to display significantly greater critical sentiment ($47.1\%$ more than expected) than citer-citee pairs that will not collaborate ($1.2\%$ more than expected; two-sided Welch's $t$-test: $t=88.4$, $p<0.0001$, $df=1003.9$). The presence of this dramatic increase could indicate that critical language in citation may be an early predictor of future collaboration.

% Fig. 4 ================================================================================================
\begin{figure*}[t]
    \centering
	\includegraphics[width=6.4 in]{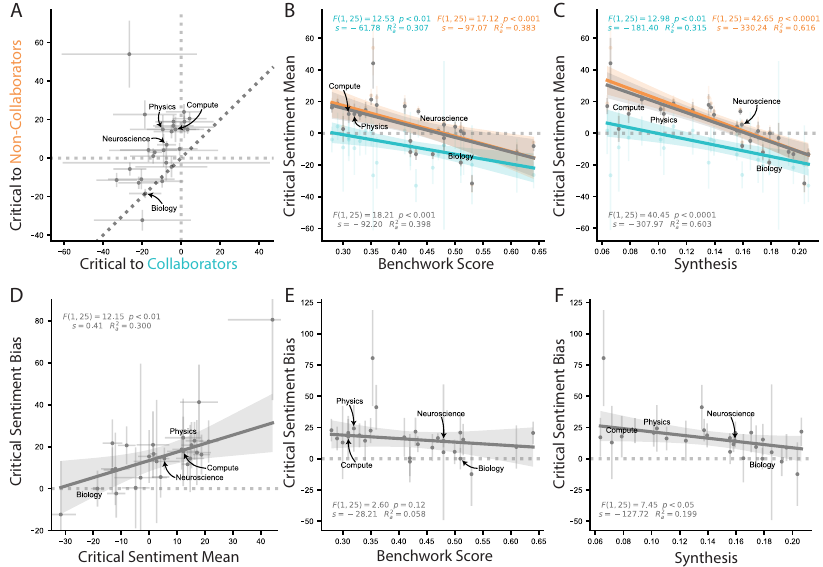}
	\caption{\textbf{Critical citation sentiment varies across disciplines.}
        \emph{(A)} Mean critical sentiment in a discipline toward collaborators versus non-collaborators.
        \emph{(B)} Mean critical sentiment as a function of the benchwork score of the citing paper's discipline. We define the benchwork score as the estimated proportion of wet lab papers (as opposed to dry lab papers) in a discipline. Orange corresponds to the sentiment toward non-collaborators, cyan to the sentiment toward collaborators, and grey to the sentiment toward both. Lines are fit using weighted least squares regression; shaded regions represent $95\%$ confidence intervals; $s$ reports the slope of the regression.
        \emph{(C)} Mean critical sentiment as a function of the synthesis score of a citing paper's discipline. We define synthesis as the proportion of review papers in a discipline.
        \emph{(D)} Relationship between the mean critical sentiment and bias. We define bias as the difference between sentiments toward non-collaborators and collaborators.
        \emph{(E)} Bias in critical sentiment as a function of a discipline's benchwork score.
        \emph{(F)} Bias in critical sentiment as a function of disciplinary synthesis.}
	\label{Fig4}
\end{figure*}

\subsection{Citation Sentiment is Most Critical from High h-index Scholars to Low h-index Non-Collaborators}

To examine how sentiment might be impacted by factors of dominance or scientific hierarchy, we consider the difference in \emph{h}-index between the citer and the citee. We first observe, regardless of the difference in \emph{h}-index, that scholars tend to express more favorable sentiment to their collaborators ($16.8\%$ more than expected) and less favorable sentiment to non-collaborators ($3.4\%$ less than expected; Fig. \ref{Fig3}A). We next observe that as citer \emph{h}-index increases relative to that of non-collaborator citees, critical sentiment increases (Fig. \ref{Fig3}B, right). These data demonstrate that citation sentiment varies with the relative status of the citer and citee.

\subsection{Citation Sentiment Tracks Disciplinary Differences \& Practices}

% Fig. 5 ================================================================================================
\begin{figure*}[t]
    \centering
	\includegraphics[width=6.4 in]{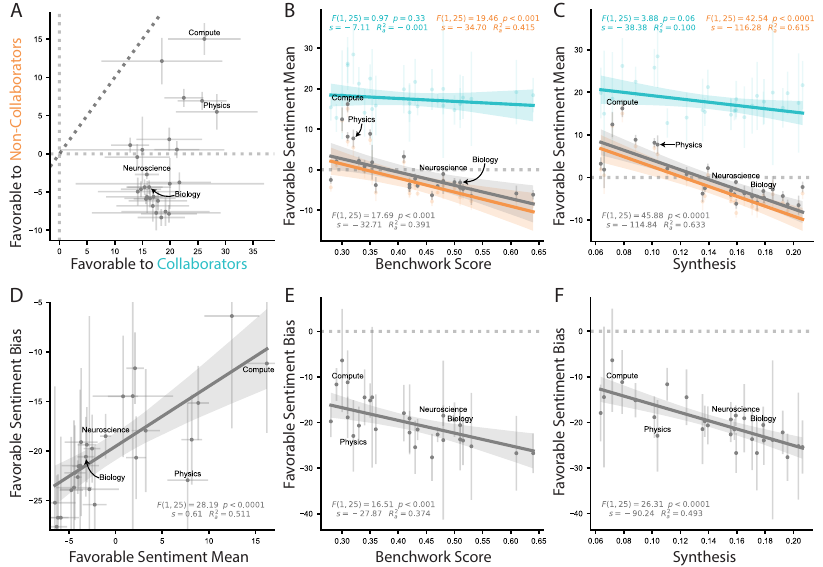}
	\caption{\textbf{Favorable citation sentiment varies across disciplines.}
        \emph{(A)} Mean favorable sentiment in a discipline toward collaborators versus non-collaborators.
        \emph{(B)} Mean favorable sentiment as a function of the benchwork score of the citing paper's discipline. We define benchwork score as the estimated proportion of wet lab papers (as opposed to dry lab papers) in a discipline. Orange corresponds to the sentiment toward non-collaborators, cyan to the sentiment toward collaborators, and grey to the sentiment toward both. Lines are fit using weighted least squares regression; shaded regions represent $95\%$ confidence intervals; $s$ reports the slope of the regression.
        \emph{(C)} Mean favorable sentiment as a function of the synthesis score of a citing paper's discipline. We define synthesis as the proportion of review papers in a discipline.
        \emph{(D)} Relationship between the mean favorable sentiment and bias. We define bias as the difference between sentiments toward non-collaborators and collaborators.
        \emph{(E)} Bias in favorable sentiment as a function of a discipline's benchwork score.
        \emph{(F)} Bias in favorable sentiment as a function of disciplinary synthesis.}
	\label{Fig5}
\end{figure*}

The previous two sections suggest that citation sentiment tracks social factors at the level of inter-personal differences (h-index) and relationships (collaboration). We expand out from dyads to larger social groups in the form of scientific disciplines, with their own broader social norms, cultures \cite{huber1990disciplinary,becher1981towards}, and identities \cite{valimaa1998culture}. Moreover, as so-called \emph{communication complexes} \cite{stichweh1992sociology} with their own rhetoric \cite{pinch1990culture}, writing styles \cite{hagge1997disciplinary}, and communicative processes \cite{stichweh1992sociology}, we expect disciplines to differ in citation sentiment and in the prevalence of critical vs. favorable sentiment to collaborators and strangers. Consistent with our expectations, we see that critical sentiment varies by discipline (Fig. \ref{Fig4}A), with most disciplines displaying less critical sentiment to collaborators than to non-collaborators (Fig. \ref{Fig4}A, the sector above the diagonal unity line). Interestingly, we observe that the degree of critical sentiment in a discipline overall is positively correlated with sentiment bias; that is, the more critical the discipline in general, the more likely the discipline will more critically cite non-collaborators than collaborators (Fig. \ref{Fig4}D; linear regression slope $s=0.41$, $95\%$ CI $(0.17,0.65)$, adjusted $R^2=0.300$; one-sided F-test $F(1,25)=12.15$, $p<0.01$; $n=27$; all disciplinary results except for brilliance result share the same sample size). Similarly, we see that favorable sentiment varies by discipline (Fig. \ref{Fig5}A), with all disciplines displaying more favorable sentiment to collaborators than to non-collaborators (Fig. \ref{Fig5}A, the sector below the diagonal unity line). The degree of favorable sentiment in a discipline overall is significantly correlated with sentiment bias (Fig. \ref{Fig5}D; slope $s=0.61$, $95\%$ CI $(0.37,0.84)$, adjusted $R^2=0.511$, $F(1,25)=28.19$, $p<0.0001$). 

\noindent \textbf{Disciplinary Differences.} Next we turn to the question of why we might expect some disciplines to be more or less critical than others, and which ones. First, we hypothesize that differences in citation sentiment will track with the differently valued explanations that scientific disciplines provide. In modern science, distinct disciplines are perceived as describing different levels or scales of nature \cite{oppenheim1958unity,potochnik2012limitations} and thereby providing differently valued explanations, with lower-level explanations being more valued than higher-level explanations (e.g., see \cite{sober1999multiple} but also \cite{ross2024causation}). This value structure is consistent with the emphasis on uncovering underlying mechanisms \cite{sep-science-mechanisms} and the prevalence of reductionism in modern science \cite{sep-scientific-reduction}. To measure a discipline's typical level of explanation, we evaluate the degree to which the discipline uses benchwork and wetlab practices, or hands-on experimentation with biological materials, chemicals, and liquids (see Supplement). Disciplines scoring high on this factor include those that provide genetic and molecular explanations, whereas disciplines scoring low on this factor include those that provide higher-level explanations from drylab practices (e.g., computational and theoretical techniques), mental processes (e.g., psychology), and social phenomena (e.g., psychiatry). 

We expect disciplines offering lower-level explanations from benchwork to have less critical sentiment because that scale of explanation is more accepted and valued. By contrast, we expect disciplines offering higher-level explanations to more frequently resort to the use of sentiment to signal valuation of the scientific work. Consistent with our expectations, we observe a significant negative trend between critical sentiment and the benchwork score (linear regression slope $s=-92.20$, $95\%$ CI $(-136.70,-47.70)$, adjusted $R^2=0.398$, $F(1,25)=18.21$, $p<0.001$; Fig. \ref{Fig4}B, grey). Interestingly, the effect holds both when citing non-collaborators ($s=-97.07$, $95\%$ CI $(-145.39,-48.75)$, $R^2=0.383$, $F(1,25)=17.12$, $p<0.001$) and when citing collaborators ($s=-61.78$, $95\%$ CI $(-97.73,-25.84)$, $R^2=0.307$, $F(1,25)=12.53$, $p<0.01$). Similarly, we observe a significant negative trend between favorable sentiment and the benchwork score (linear regression slope $s=-32.71$, $95\%$ CI $(-48.72,-16.69)$, adjusted $R^2=0.391$; one-sided F-test $F(1,25)=17.69$, $p<0.001$; Fig. \ref{Fig5}B, grey). Again the effect holds when citing non-collaborators ($s=-34.7$, $95\%$ CI $(-50.90,-18.50)$, $R^2=0.415$, $F(1,25)=19.46$, $p<0.001$), but is significantly mitigated when citing collaborators ($R^2=-0.001$, $F(1,25)=0.97$, $p=0.33$). Further, we observe a significantly negative trend between the benchwork score and bias in favorable sentiment towards collaborators (slope $s=-27.87$, $95\%$ CI $(-42.00,-13.74)$, adjusted $R^2=0.374$, $F(1,25)=16.51$, $p<0.001$; Fig. \ref{Fig5}E). This negative trend is not seen for critical sentiment (linear fit and one-sided F-test between benchwork score and critical sentiment bias $R^2=0.058$, $F(1,25)=2.60$, $p=0.12$; Fig. \ref{Fig4}E).

\noindent \textbf{Disciplinary Practices.} The previous section provides evidence that citation sentiment tracks disciplinary differences in the level of scientific explanation. But apart from a discipline's nature, might its practices also play a role? If a discipline engages in inclusive communication practices---for example, by regularly publishing review articles that synthesize the work of different groups, computational methods, experimental approaches, and theoretical perspectives---we might expect its scholars to be exposed to a greater diversity of ideas, appreciate the value in diverse efforts, and cite less critically. This expectation is supported by work in communication and polarization. Exposure to like-minded information can have a polarizing effect \cite{garrett2014implications}, whereas exposure to multiple competing views can increase tolerance and familiarity \cite{price2002does}, accompany intellectual humility \cite{porter2017intellectual}, and decrement polarization \cite{sunstein2007repulic}. In providing exposure to diverse views, scientific review articles could further attenuate polarization by modeling ingroup open-mindedness \cite{wojcieszak2020social}: a receptiveness to new ideas, a capacity to embrace a variety of views, and an expanded `latitude of acceptance', or the range of views that one finds to be acceptable for other people to hold \cite{baehr2011structure,christensen2019reopening,dieffenbach2023belief}. Accordingly, we expect that disciplines with fewer review articles---a more fragmented, heterogeneous, and potentially polarized scholarly landscape---would evince more critical sentiment. 

We define synthesis to be the proportion of review papers in a discipline. Consistent with our expectations, we observe a significant negative correlation between critical sentiment and synthesis (linear regression slope $s=-307.97$, $95\%$ CI $(-407.70,-208.23)$, adjusted $R^2=0.603$; one-sided F-test $F(1,25)=40.45$, $p<0.0001$; Fig. \ref{Fig4}C, grey). The effect holds when citing non-collaborators ($s=-330.24$, $95\%$ CI $(-434.39,-226.10)$, $R^2=0.616$, $F(1,25)=42.65$, $p<0.0001$), but is somewhat mitigated when citing collaborators ($s=-181.40$, $95\%$ CI $(-285.11,-77.70)$, $R^2=0.315$, $F(1,25)=12.98$, $p<0.01$). Further, we observe that disciplines with greater synthesis display less bias in critical sentiment towards non-collaborators, and disciplines with less synthesis display more bias (linear fit and one-sided F-test between synthesis and critical sentiment bias $s=-127.72$, $95\%$ CI $(-224.11,-31.33)$, $R^2=0.199$, $F(1,25)=7.45$, $p<0.05$; Fig. \ref{Fig4}F). Similarly, we observe a significant negative correlation between favorable sentiment and synthesis (linear regression slope $s=-114.84$, $95\%$ CI $(-149.75,-79.92)$, adjusted $R^2=0.633$; one-sided F-test $F(1,25)=45.88$, $p<0.0001$; Fig. \ref{Fig5}C, grey). Again the effect holds when citing non-collaborators ($s=-116.28$, $95\%$ CI $(-153.00,-79.56)$, $R^2=0.615$, $F(1,25)=42.54$, $p<0.0001$), and is mitigated when citing collaborators ($R^2=0.100$, $F(1,25)=3.88$, $p=0.06$). We observe a significant negative relationship between synthesis and bias in favorable sentiment towards collaborators (slope $s=-90.24$, $95\%$ CI $(-126.48,-54.01)$, adjusted $R^2=0.493$, $F(1,25)=26.31$, $p<0.0001$; Fig. \ref{Fig5}F). Note that our null model (see Methods) accounts for the fact that review papers in general express less sentiment ($35.64\%$ favorable, $6.79\%$ critical) compared to research papers ($44.71\%$ favorable, $8.37\%$ critical).
This result also holds before the sentiment aggregation step in our methods (see Supplement).

\noindent \textbf{Specificity Analyses.} Collectively, these results provide evidence that citation sentiment is associated with disciplinary differences in the level of explanation and disciplinary practices of review and synthesis. In the supplementary materials, we assess the specificity of our findings by also examining disciplinary differences in perceived brilliance \cite{leslie2015expectations,muradoglu2022women}.

\subsection{Citation Sentiment Tracks Country Differences in Individualism and Power}

% Fig. 6 ================================================================================================
\begin{figure*}[t]
    \centering
	\includegraphics[width=6.4 in]{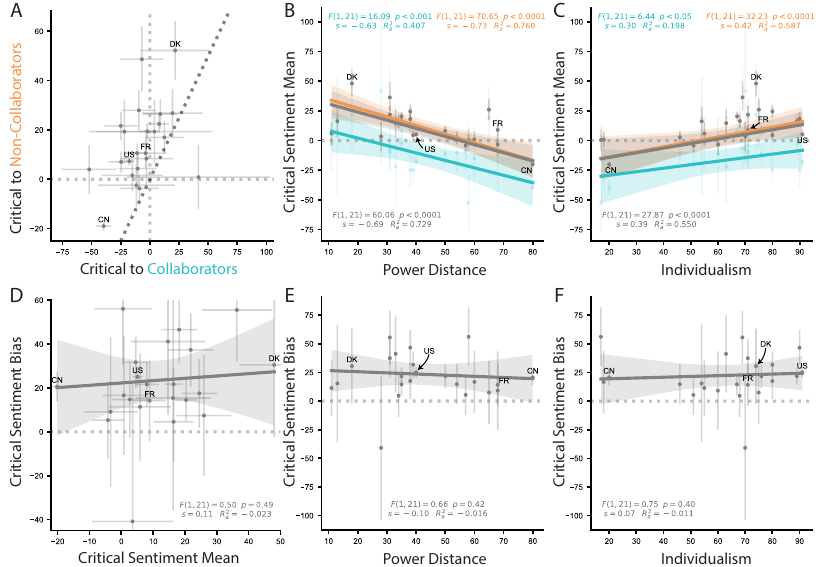}
	\caption{\textbf{Critical sentiment varies across countries.}
        \emph{(A)} Mean critical sentiment across countries toward collaborators versus non-collaborators.
        \emph{(B)} Mean critical sentiment for countries as a function of the degree to which people accept the unequal distribution of power, measured as the power-distance \cite{hofstede2001culture}.
        \emph{(C)} Mean critical sentiment for countries as a function of individualism; individualism measures the degree to which people believe that the interests of the individual should take precedence over the interests of a social group \cite{hofstede2001culture}.
        \emph{(D)} Relationship between the mean and bias in critical sentiment across countries.
        \emph{(E)} Bias in critical sentiment across countries as a function of power-distance.
        \emph{(F)} Bias in critical sentiment across countries as a function of individualism.}
	\label{Fig6}
\end{figure*}

\begin{figure*}[t]
    \centering
	\includegraphics[width=6.4 in]{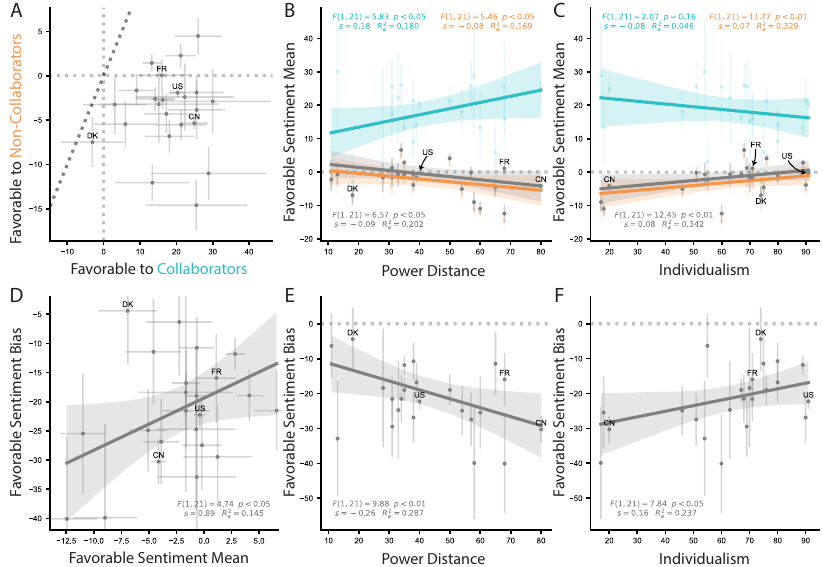}
	\caption{\textbf{Favorable sentiment varies across countries.}
        \emph{(A)} Mean favorable sentiment across countries toward collaborators versus non-collaborators.
        \emph{(B)} Mean favorable sentiment for countries as a function of the degree to which people accept the unequal distribution of power, measured as the power-distance \cite{hofstede2001culture}.
        \emph{(C)} Mean favorable sentiment for countries as a function of individualism; individualism measures the degree to which people believe that the interests of the individual should take precedence over the interests of a social group \cite{hofstede2001culture}.
        \emph{(D)} Relationship between the mean and bias in favorable sentiment across countries.
        \emph{(E)} Bias in favorable sentiment across countries as a function of power-distance.
        \emph{(F)} Bias in favorable sentiment across countries as a function of individualism.}
	\label{Fig7}
\end{figure*}

In the previous sections, we considered a scholar's location within a range of social structures from small collaborations to wider disciplines. But the sociocultural milieu in which each scholar exists also extends outwards to the national scale, with each country having distinct norms that can manifest in scientific practice. Cultural differences in academic writing are well-known, and typically studied under the framing of intercultural or contrastive rhetoric and applied linguistics \cite{siepmann2006academic,kubota2010cross,atkinson2020intercultural}. Importantly, smaller cultures (e.g., personal, disciplinary) can interact with national culture \cite{atkinson2004contrasting}, often in more-or-less tacit and unthinking writing practices \cite{atkinson2003writing}. Prior work has examined cultural differences in economic functionality, capitalism, and individualism that manifest in both academic and popular writing  \cite{atkinson2003writing,ramanathan1999individualism,guntuku2023historical}. Importantly, cultural differences extend beyond writing style to citation practices \cite{hu2014disciplinary}. For example, citations differ across cultures in voice and stance \cite{flottum2006academic,flottum2012variation}, density \cite{taylor1991linguistic}, dialogic contraction/expansion \cite{hu2014disciplinary}, and acknowledgment of limitations \cite{sun2024results}. Here we show that cultural differences also exist in citation sentiment (Fig. \ref{Fig6}A and Fig. \ref{Fig7}A), and seek to understand how variations in citation sentiment may reflect cultural norms around individualism and acceptance of the hierarchical structures of society. 

Accordingly, we examine power-distance and individualism---two key dimensions of modern culture \cite{hofstede2001culture,hofstede2015geert}---to assess country-level differences in citation sentiment. Power-distance refers to the degree to which people accept the unequal distribution of power. Individualism refers to the degree to which people believe that the interests of the individual should have precedence over the interests of a social group, and to which people oppose external interference upon individual interests by society or institutions such as the government \cite{hofstede2001culture}. We observe a significant negative correlation between critical sentiment and power-distance (Fig. \ref{Fig6}B; linear regression slope $s=-0.69$, $95\%$ CI $(-0.87,-0.50)$, adjusted $R^2=0.729$; one-sided F-test $F(1,21)=60.06$, $p<0.0001$; $n=23$; all country-level results share the same sample size), such that countries whose people accept the unequal distribution of power tend to evince less critical citation sentiment. The effect holds for both non-collaborators ($s=-0.73$, $95\%$ CI $(-0.91,-0.55)$, $R^2=0.760$, $F(1,21)=70.65$, $p<0.0001$) and collaborators ($s=-0.63$, $95\%$ CI $(-0.95,-0.30)$, $R^2=0.407$, $F(1,21)=16.09$, $p<0.001$). Further, we observe a significant positive trend between critical sentiment and individualism (Fig. \ref{Fig6}C; linear regression slope $s=0.39$, $95\%$ CI $(0.24,0.54)$, adjusted $R^2=0.550$; one-sided F-test $F(1,21)=27.87$, $p<0.0001$), such that countries whose people privilege individual interests over group interests tend to evince more critical citation sentiment. This effect again holds for both non-collaborators ($s=0.42$, $95\%$ CI $(0.26,0.57)$, $R^2=0.587$, $F(1,21)=32.23$, $p<0.0001$) and collaborators ($s=0.30$, $95\%$ CI $(0.05,0.54)$, $R^2=0.198$, one-sided F-test $F(1,21)=6.44$, $p<0.05$).
The overall trends between favorable sentiment and power-distance (Fig. \ref{Fig7}B, grey) or individualism (Fig. \ref{Fig7}C, grey) are similar but weaker in size compared to those of critical sentiment.
Interestingly, unlike critical sentiment, an opposite trend appears between favorable sentiment towards collaborators and power-distance (Fig. \ref{Fig7}B, cyan, slope $s=0.18$, $95\%$ CI $(0.03,0.34)$, adjusted $R^2=0.180$, one-sided F-test $F(1,21)=5.83$, $p<0.05$).
The bias in critical sentiment towards non-collaborators as opposed to collaborators (Fig. \ref{Fig6}D) does not track individualism or power-distance (Fig. \ref{Fig6}E-F), but favorable sentiment does (Fig. \ref{Fig7}E, slope $s=-0.26$, $95\%$ CI $(-0.42,-0.09)$, adjusted $R^2=0.287$, one-sided F-test $F(1,21)=9.88$, $p<0.01$ for power-distance; Fig. \ref{Fig7}F, slope $s=0.16$, $95\%$ CI $(0.04,0.28)$, adjusted $R^2=0.237$, one-sided F-test $F(1,21)=7.84$, $p<0.05$ for individualism).
We observe a weak trend between average favorable sentiment expression and the bias towards collaborators (Fig. \ref{Fig7}D).
These results suggest that norms around social hierarchy may influence citation sentiment, resulting in a stronger expression of differences and criticisms in low-power and high-individualism cultures.

% Fig. 7 ================================================================================================

% Fig. 8 ================================================================================================
\subsection{Citation Sentiment and Gender}
\begin{figure*}[t]
    \centering
	\includegraphics[width=6.4 in]{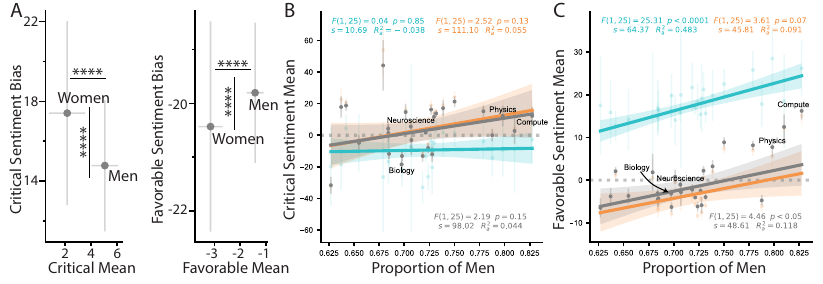}
	\caption{\textbf{Citation sentiment varies by gender.}
        \emph{(A)} Mean and bias in citation sentiment separated by sentiment type and author gender.
        Asterisks indicate $p$-values in two-sided Welch's $t$-tests; a single asterisk indicates $p<0.05$; two asterisks indicate $p<0.01$; three asterisks indicate $p<0.001$; and four asterisks indicate $p<0.0001$.
        \emph{(B)} Mean critical sentiment for disciplines as a function of the proportion of men.
        \emph{(C)} Mean favorable sentiment for disciplines as a function of the proportion of men.}
	\label{Fig8}
\end{figure*}

Thus far, the data consistently suggest that citation sentiment can track ingroup/outgroup relations, structures of dominance, and hierarchies that are interpersonal (collaboration), prestigious (h-index, disciplinary level of explanation), and national (beliefs about power and society). Yet, an element that can influence all of these factors is personal identity; dimensions of identity such as gender, sex, race, ethnicity, class, and (dis)ability determine one's placement, status, and power in social groups across scales. How might personal identity manifest in citation sentiment, again marking structures of dominance and hierarchy? To address this question, we examined author gender, which has a known relation to writing style (including language use, rhetorical strategies, and authorial positioning \cite{deats1994gender,ghafoori2012comparative,robson2002writes,sala2009gender,novell2001gendered}) and citation practice \cite{lariviere2013bibliometrics,chatterjee2021gender,dworkin2020extent,fulvio2021gender,teich2022citation,caplar2017quantitative}.

Compared to the null model, we observe that men display $2.89\%$ more critical sentiment than women (Fig. \ref{Fig8}A; two-sided Welch's t-test: $t=52.55$, $p<0.0001$, $df=1851$) and $1.72\%$ more favorable sentiment than woman (Fig. \ref{Fig8}A; two-sided Welch's t-test: $t=95.40$, $p<0.0001$, $df=1787$). This pattern of results indicates that men write with more sentiment overall than women. Interestingly, men also display $0.63\%$ less favorable sentiment bias than women (two-sided Welch's t-test: $t=8.46$, $p<0.0001$, $df=1741$; Fig. \ref{Fig8}A), and $2.63\%$ less critical sentiment bias than women (two-sided Welch's t-test: $t=-14.66$, $p<0.0001$, $df=1808$). This pattern of results indicates that men more evenly distribute their sentiment across both collaborators and non-collaborators compared to women. 

Because gender ratios vary by discipline \cite{nsf2023stats,cheryan2017why,leslie2015expectations}, we next examined whether citation sentiment tracked the proportion of men in a discipline. We observe that the proportion of men is not correlated with critical sentiment (Fig. \ref{Fig8}B; linear regression adjusted $R^2=0.044$; one-sided F-test $F(1,25)=2.19$, $p=0.15$) but it is positively correlated with favorable sentiment towards collaborators (Fig. \ref{Fig8}C; slope $s=64.37$, $95\%$ CI $(38.02,90.72)$, adjusted $R^2=0.483$; one-sided F-test $F(1,25)=25.31$, $p<0.0001$) and towards others in general (Fig. \ref{Fig8}C; slope $s=48.61$, $95\%$ CI $(1.22,96.00)$, adjusted $R^2=0.118$; one-sided F-test $F(1,25)=4.46$, $p<0.05$). Generally, these results indicate that gender interacts with both discipline and collaboration status in explaining citation sentiment.

\section{Discussion}

Scientific writing is a social act \cite{bazerman1983scientific}, a locus of cultural production \cite{medin2014cultural}, and a means of communicating to our peers with a variety of purposes \cite{goldbort2006writing}. A key communication channel in scientific writing is citations, and the sentiments that surround them. Early work examining this channel assessed citations that carried sentiments of evaluation and negation by cataloging the verbs used in citation sentences \cite{thompson1991evaluation}. Later studies employed computational methods to assess the functions and motivations of citations based on their content \cite{ding2014content}. Those functions have been summarized in terms of positive, negative, and neutral sentiment \cite{teufel2006automatic}, as well as in terms of uncertainty versus utility \cite{small2011interpreting}. Here we significantly extend prior work by expanding the size of the database and employing a large language model to efficiently and fairly quantify sentiment in the citation sentences of over a hundred thousand papers across $181$ neuroscience journals. Our approach allows us to uncover significant relations between citation sentiment and sociocultural factors across scales of local collaboration, broader discipline, and whole country. Collectively, the data point to a focal role for status, hierarchy, and power in the use of sentiment in modern citation practices.

\subsection{Relations of authorship and status}

Communication is critical for cooperation \cite{noe2006cooperation}. In science, cooperation can take on many forms, but one key type that has risen in prominence recently is collaboration \cite{adams2012collaborations,wuchty2007increasing}. How might communication through citation sentiment reflect and support collaborative relationships? Here we evaluate collaboration via coauthorship---a widely accepted proxy even though not all collaborations are rewarded with coauthorship \cite{laudel2002what}. We find that scholars cite their collaborators more favorably and less critically than they cite non-collaborators. Utilizing science communication practices such as favorable sentiment could support collaboration by increasing trust, reducing competition, and supporting the social cohesion that allows for knowledge transfer \cite{fleming2007collaborative}. In turn, scholarly collaboration can enhance the quality of the scientific work \cite{costantini2010effect}, which can then provide scholars with greater status in the scientific community.

Communication is not only critical for cooperation but also for sustaining systems of social status, which themselves can either support \cite{kawakatsu2024mechanistic} or undermine cooperation \cite{chakrabarti2022status}. In science, a prominent system of social status or hierarchy combines research productivity and quality \cite{petersen2014reputation}. As a useful proxy, we employ the h-index and show that differences between the h-index of the citer and the citee do not significantly track citation sentiment among collaborators. However, when a high h-index scholar cites a lower h-index non-collaborator, they do so with less favorable and more critical sentiment. Because we do not have a measure of a scholar's seniority, we cannot determine whether this behavior arises from more senior scholars critically citing the work of similarly-senior or less-senior scholars, or from less senior scholars critically citing the work of more senior scholars with lower h-indices. The graded nature of this relationship is notable, and consistent with prior work demonstrating that as the steepness of a dominance hierarchy increases, aggression from dominants to subordinates increases while reciprocal exchanges decrease \cite{jaeggi2016obstacles}. Interestingly, this behavior tends to exist in communities with scarce resources, but is mitigated in those that depend heavily on coalition formation \cite{rueden2019dynamics,jaeggi2016obstacles}. It is possible that the success of high h-index scholars is less dependent on scientific collaboration or coalition than that of low h-index scholars, and that critical sentiment serves to underscore dominance. 

\subsection{Relations of disciplinary nature and practice}
Systems of social status extend beyond the single scholar to larger groups, subfield, and disciplines. Disciplines differ in their values, practices, criteria for progress, models, justification, and evidence, as well as their level of explanation \cite{oppenheim1958unity,potochnik2012limitations,sober1999multiple}. As such, disciplines can be differently valued, with research offering lower-level explanation and mechanisms valued as particularly powerful and foundational \cite{ross2024causation}. A key open question is how citation practices might track these disciplinary differences and markers of status. Prior work has uncovered disciplinary differences in citation density \cite{flottum2006academic,hyland1999academic,thompson2001looking,sanchez2018reference}, citation functions \cite{harwood2009interview,small2011interpreting}, citation integration \cite{flottum2006academic,hyland1999academic,thompson2001looking}, sources of citations \cite{charles2006construction,pecorari2006visible}, types and tenses of reporting clauses \cite{charles2006construction}, frequency of reporting verbs \cite{hyland1999academic}, and preferences for particular types of reporting verbs \cite{flottum2006academic,charles2006construction,hyland2002activity}. Here we extend these observations by demonstrating that citation sentiment tracts disciplinary differences in benchwork versus computational or theoretical work. We observe that disciplines offering lower-level explanations from benchwork have less sentiment overall (whether critical or favorable), consistent with the hypothesis that the scale of explanation in benchwork disciplines is more accepted and valued. We further observe that disciplines offering higher-level explanations have greater citation sentiment (both critical and favorable), which might reflect a practice of using sentiment to signal valuation of scientific work when the scale of explanation is less accepted and valued. These potential sociocultural explanations for citation sentiment align with prior work, which interprets disciplinary differences as influenced by the epistemologies underlying a discipline's cultural practices and the ethnolinguistic norms of disciplinary communication \cite{hu2014disciplinary}. 

Beyond inherent level of explanation and type of knowledge, a discipline can also be characterized by its more practical methods and practices \cite{lenoir1997discipline}. Here we were interested in practices of scholarly publication that might modulate the social structure of a discipline---including the degree of divisiveness or fragmentation---and hence that could change the balance between critical versus favorable sentiment in citations. In particular, we investigated the markedly growing practice of publishing review articles \cite{chandra2024shifting}, and found that disciplines which published a larger ratio of review articles to empirical articles displayed less sentiment overall (either critical or favorable), instead citing science more neutrally and non-affectively. It is interesting to consider this finding in light of prior work on motivated reasoning \cite{kunda1990case}, which can influence the processing of scientific evidence by subconsciously skewing it in a way that supports an individual's prior beliefs \cite{stenhouse2018potential}. Scientific review articles could serve as interventions for open-mindedness \cite{dolbier2024open}, for example by puncturing the so-called illusion of explanatory depth (whereby people think that they know more about complex phenomena than they really do) \cite{rozenblit2002misunderstood}, providing exposure to the perspectives of scholars with other viewpoints \cite{todd2014perspective}, and foregrounding incremental theory, in the sense that knowledge is changeable and unlimited, as opposed to entity theory, where knowledge is fixed or finite \cite{dweck2012implicit}.

\subsection{Relations of culture across countries}
Although scientists live and work within an explicitly scholarly social milieu---comprised of collaborative relationships and disciplinary identities---science and its practitioners also exist within other cultural structures outside of the academy. For example, research priorities and funding levels differ significantly across countries \cite{confraria2024countries,petersen2021inequality}. Further, social structures of prestige and marginalization impact what science is done and who gets to do it \cite{kozlowski2022intersectional,morgan2018prestige}, often with detrimental outcomes for discovery and innovation \cite{mongeon2016concentration,hofstra2020diversity}. Here we asked whether the well-known cultural differences in academic writing extend to citation practices, and whether the sentiment of those citations tracks cultural norms around status, hierarchy, and power. We found that countries characterized by individualism displayed greater citation sentiment (both critical and favorable) to non-collaborators, indicating a more pervasive use of affect in scientific writing. By contrast, countries characterized by collectivism displayed less citation sentiment, utilizing a more neutral writing style. We further found that countries whose people tended to accept the unequal distribution of power (as quantified by the power-distance) displayed less citation sentiment (both critical and favorable) to non-collaborators. It is possible that citation sentiment is used in science to uphold or dismantle existing structures of status, hierarchy, and dominance whereas the lack of sentiment indicates an acceptance of power and a focus on collective goals. More broadly, our results suggest that the non-scholarly cultural systems that we as scientists are embedded within play a key role in how scholars practice science individually.   

\subsection{Gender differences in citation sentiment}

The non-scholarly systems that scientists work within exist not only at the large scale of countries but also at the small scale of identities. A key sociocultural difference exists between those people socialized (or living) as women and those socialized (or living) as men \cite{leaper2007socialization}. These differences arise along multiple dimensions of experience that could impact science, for example in linguistic behavior \cite{coates2015sociolinguistic} and in social behavior \cite{hyde2014gender}. Consistent with these general observations, gender differences have been identified in how scholars engage in the scientific process, from writing styles \cite{dejesus2021when} and research areas \cite{kim2022gendered} to self-promotion practices \cite{peng2023gender}, self-citation \cite{king2017men}, positive presentation of their own results \cite{lerchenmueller2019gender}, and use generic statements, framing an idea as broad, timeless, and universally true \cite{dejesus2021when}. Here we demonstrate that those gender differences also extend to the sentiment with which scholars cite one another. We find that men cite with greater sentiment---both critical and favorable---and women cite with greater sentiment bias, being more favorable to collaborators than to non-collaborators. 

Why might men cite with more affect than women? Prior work has provided evidence that men engage socially in science, more so than women; for example, men are more likely to share their papers socially, and preferentially do so with other men \cite{massen2017sharing}. This out-sized social engagement in science holds both in laboratory settings and in evaluations of real-world behavior on social media \cite{zhu2019gender,peng2023gender}. Broadly, it suggests that men may treat science as a small-scale society, where prior evidence marks a key interdependence between cooperation (here perhaps supported by favorable sentiment) and status hierarchy (here perhaps upheld by critical sentiment) \cite{rueden2019dynamics}. Why might women cite with greater sentiment bias than men, assigning more favorable (and less critical) sentiment to collaborators than to non-collaborators? Prior work suggests that women, on average, tend to use more affiliative language whereas men use more aggressive language (and are more generally talkative) \cite{leaper2007meta}. Those affiliative tendencies may manifest in the use of citation sentiment to strengthen in-group relations, via more favorable and less critical language. Future work could consider how these gender differences might intersect with other dimensions of identity including race, ethnicity, sexual orientation, and socioeconomic status.

\subsection{Methodological Considerations}

Several methodological considerations are pertinent to our study. First, our sentiment analysis method captures both evaluative judgments and expressed commonalities and differences between the citing and the cited works. Sentiment can be defined in multiple ways \cite{liu2017many}, and the findings must be interpreted in the context of the definition used. Second, because we have limited data for each individual author, we cannot predict how a single citer might behave. Instead, our results characterize trends for a group or groups of people. Third, in seeking to understand how scholars cited the work of others with different sorts of affect, it was important to minimize the effect of content. Hence, we developed and employed a null model that accounts for textual (dis)similarity between citing and cited papers, and show residual sentiment estimates. Fourth, studies of culture are challenged by the complexity of the concept, and prior work has underscored the need to think about multiple dimensions and scales of culture \cite{atkinson2004contrasting,atkinson2003writing}. Here we acknowledge that complexity by examining cultural factors that span from small to large scales, and concretize our investigation by using specific variables related to social practices and beliefs. Fifth, prior work has found that criticism and intellectual differences in citations are rare (from $0.8\%$ \cite{nicholson2021scite} to $15\%$ \cite{yan2020relationship}). To ensure a reliable signal, we employed a large database of $627108$ citation sentences and a powerful and sensitive sentiment detection method using a large language model. Sixth, our results are limited to the field of neuroscience, and it is not yet clear whether they are pertinent to other fields. It seems likely that there could be links to other young disciplines, that have common factors like a relative youth which could allow for the emergence of still-malleable subdisciplines, a marked interdisciplinarity drawing on older fields, and a diversity of opinions, methods, and experimental approaches \cite{sabbatini2002interdisciplinarity,bilyk2022neurobiology}. It would be interesting to contrast this work to other disciplines of varying ages.

\section{Conclusion}
Growing evidence across the fields of psychology, sociology, and science of science underscores the humanity that scholars bring to the scientific enterprise. Individually, we bring our epistemological standpoints, cognitive biases, perceptions, and values whereas in groups we bring disciplinary norms of preferred practices, methods, standards, and explanations. Here we uncover human subjectivities in scientific citation, demonstrating that citation sentiment in neuroscience tracks multiscale sociocultural norms of status, collaboration, discipline, and country. Uncovering the social construction of science is not only a scholarly contribution but also an ethical one, as obscuring that construction is an epistemic harm \cite{hall2017obscured}. Looking to the future, important open questions remain, including how citation sentiment might have changed over time, how the relation between sentiment and culture might be impacted by serendipitous discoveries or paradigm shifts, and how other aspects of affect might manifest in science communication more broadly.

\section{Methods}

\subsection{Citation and Author Data}

We used PubMed Central and the Web of Science API to build our datasets for analyses.
The PubMed Central (PMC) Open Access Subset (OA) includes millions of research articles and pre-prints alongside relevant metadata, such as article type, venue of publication, and institutional information \cite{pmcoas}. 
We gathered all English language research and review articles from PMC OA relating to neuroscience that were published between $1998$ and $2023$ and that cited at least one article.
In parallel, using the Web of Science, we compiled a list of journals classified under the ``neurosciences'' category with an impact factor of at least $3$ in the year $2022$.
We manually added three more journals creating a list of $181$ journals in which the gathered articles were published; the full list is available in the Supplement.
Each article in the PMC OA subset was stored in an extensible markup language (XML) file with attached tags delineating its several components.
For instance, the \texttt{<article-title>} tag encloses an article's title, \texttt{<abstract>} encloses the abstract, and \texttt{<xref>} encloses a citation in the body of the text.
Using these and other relevant tags, we extracted each article's title, list of authors, affiliations (at the departmental, institutional, and country levels), and all sentences containing citations to other works in the same dataset.
We obtained $69.24\%$ of last authors' h-indices from the Web of Science.
We used Gender API to infer gender from the names of the last authors, requiring the underlying model to be at least $70$\% confident in its prediction \cite{genderapi}.
The resulting dataset contained $627108$ citation sentences, $383292$ unique authors, $56034$ unique last authors; from the data, $27$ departments and $23$ countries and regions were included in the analyses.
Additionally, we built a collaboration network where nodes correspond to authors and edges correspond to co-authorship.
From this network, we define collaboration distance as the length of the shortest path between two individuals.

\subsection{Measuring Sentiment}

We assessed citation sentiment using \texttt{GPT-3.5-Turbo-1106}, a large language model (LLM) by OpenAI.
LLMs are trained on a large corpus of text data and can generate language across various domains.
Through their exposure during training to diverse contexts, they can recognize subjective elements of human language, such as tone \cite{zhang-etal-2024-sentiment}, irony \cite{Tomas2023, irony_2023}, sarcasm \cite{sarcasm_RAG_2024}, and implied meaning \cite{implied_meaning_2024}.
Here, we prompted \texttt{GPT-3.5-Turbo-1106} to evaluate a citation as negative, neutral, or positive; prompt text supplied to the LLM is available in the Supplement.
Each entry in the citation data contained a single sentence with possibly multiple citations.
Consider, for instance, the following sentence with two citations: ``Lorem ipsum odor amet [citation 1], consectetuer adipiscing elit [citation 2].''
This text results in two entries for the LLM to judge: (1) ``Lorem ipsum odor amet [citation 1], consectetuer adipiscing elit'', and (2) ``Lorem ipsum odor amet, consectetuer adipiscing elit [citation 2].''
Using the OpenAI API, we passed each entry to the model separately for sentiment analysis.
In practice, each citation location in a sentence was replaced by the Unicode character with encoding U+272A (see Supplement for full model prompt).
It is common for an article to cite another article multiple times in different locations in the text.
We aggregated overall sentiment for a given citer-citee pair through the following order of precedence: negative sentiment took precedence over positive sentiment, and both took precedence over neutral sentiment.
Here, we assumed negative sentiment from a citing article overrides any positive or neutral citations \cite{Catalini_2015, Bordignon2020}.
We used aggregated sentiment throughout the analysis.

However, before using the model in this manner, we performed an additional fine-tuning step.
LLMs can perform various tasks through zero-shot prompting, meaning they do not generally require more training.
Supplying a well-chosen prompt is often sufficient.
However, a small amount of domain-specific fine-tuning can improve performance and help generate well-formatted outputs for easier parsing.
The fine-tuning process we used entailed adapting a large language model trained for a general use case to a task-specific use case, in this case, assessing citation sentiment.
With the help of $5$ volunteer researchers, we manually annotated $300$ randomly chosen citations as negative ($-1$), neutral ($0$), or positive ($1$).
These citations were drawn from journals in neuroscience and physics, the former high in critical sentiment and the latter low in critical sentiment \cite{lamers2021meta}, covering a broad range of sentiment expression.
The fine-tuning dataset comprised these citations and their annotator-assigned labels.
If the average sentiment value was less than or equal to $-0.4$, we assigned a negative label; if it is greater than or equal to $0.4$, we assigned a positive label.
All citations with values between the two were assumed to be neutral.
Fine-tuning \texttt{GPT-3.5-Turbo-1106} with this data resulted in an improvement of $28\%$ over the baseline (see Supplement).

We used a null model to benchmark observed sentiment values against those that might emerge by random chance.
For instance, we found that papers with similar content were likely to involve negative citations, thereby increasing the likelihood that such a citation in the data was an outcome of the similarity of work and not an underlying bias of interest.
Similar effects were associated with article type (review vs. research) and the number of times an article references another.
A review article tends to be more neutral in sentiment compared to a research article, and an article that cites another article more frequently tends to be more negative. 
We used these factors to measure the expected probability of observing a negative, neutral, or positive sentiment for any given citer-citee pair.
First, relating to shared content, we created three bins for high, medium, and low similarity, where similarity was measured as the cosine distance between the vector embeddings of article titles.
The article titles were extracted using a PMC parser \cite{Achakulvisut2020}.
The embeddings were generated using OpenAI's \texttt{text-embedding-3-small} model.
Additionally, $15$ bins were associated with citation frequency and $2$ with article type.
Initially, for each of the $90$ ($3 \times 15 \times 2$) resulting combinations, we performed maximum likelihood estimation to compute the probability that a citation in the data belonged to a given group.
However, if a bin had fewer than $500$ samples, we used the marginal distribution over title similarity to create a larger bin.
In these instances, we chose to ignore the impact of title similarity since it had the weakest effect on citation valence compared to article type and citation frequency.
Further, when a citing-cited pair had a citation frequency greater than $15$, we used the corresponding bin for $15$ citations instead.
We used the probabilities underlying the $54$ resulting bins to draw realizations of null sentiment values for given citing-cited pairs.
These realizations served as the basis of comparison throughout our analyses.

Consider a set of citations corresponding to a broader category of interest, such as citations between collaborators or citations between authors with an h-index difference of $2$.
Given such a set, we can measure the ratio of positive, neutral, and negative citations within it.
However, if the set is small in size, the estimated ratios may not accurately represent the true underlying distribution of sentiments for the group.
To mitigate this concern, we applied bootstrapping.
We created $1000$ resampled sets by randomly sampling from the original, each time with replacement, thereby allowing us to estimate a distribution of ratios for each type of sentiment.
Bootstrapping in this manner helps to construct robust distributions and derive more accurate measures of variability.
Simultaneously, for each of the $1000$ resampled sets for a given sentiment type, we evaluated the expected ratios considering the null model.
Our measure of interest was the percentage difference in the empirical ratio relative to the ratio derived from the null model.
In all figures, we plot mean values for the normalized empirical ratios and estimate error bars using standard deviations.

\subsection{Departmental and Cultural Measures}

Alongside citation sentiment, we evaluated other measures at the broader departmental and cultural scales.
To examine the disciplinary impact on sentiment, we selected $28$ departments (see Supplement for more details).
For the corresponding regression analyses of disciplinary and global scales, we included only the subsets that have at least $100$ citations towards collaborators to ensure sufficient sample sizes.
This procedure resulted in $27$ departments and $23$ countries and regions.
At the departmental level, we evaluated the proportion of articles originating in benchwork, terming this measure the \textit{benchwork score}.
Lower values corresponded to computational settings, whereas larger values corresponded to research performed predominantly in wet labs.
We measured this score using a base version (un-finetuned) of \texttt{GPT-3.5-Turbo-0125}.
For each of the $28$ departments, we first sampled $100$ articles, excluding reviews.
We then prompted the model to determine whether a given article was a product of benchwork by examining most of its content.
The content was extracted using the same PMC parser \cite{Achakulvisut2020} as the one for title extraction, and up to $10000$ tokens worth of text were included for each article.
The benchwork score for a department was defined as the fraction of papers identified by the model having originated in a wet lab.
The prompt text supplied to the LLM is available in the Supplement.
We also measured \textit{synthesis} as the proportion of review papers among all papers in a discipline.
We identified a review paper by the \texttt{article-type} variable from the XML file for each paper.
At the broader cultural scale, we used the 6-D model of national culture to measure power-distance and individualism.
Power-distance is the ``extent to which the less powerful members of organizations and institutions (like the family) accept and expect that power is distributed unequally.''
Individualism, in turn, is the ``extent to which people feel independent, as opposed to being interdependent as members of larger wholes.''
Values for these measures are available at \cite{hofstede2015geert,hofstede2001culture}.

\section{Citation Diversity Statement}
Recent work in several fields of science has identified a bias in citation practices such that papers from women and other minority scholars are under-cited relative to the number of such papers in the field \cite{mitchell2013gendered,dion2018gendered,caplar2017quantitative, maliniak2013gender, Dworkin2020.01.03.894378, bertolero2021racial, wang2021gendered, chatterjee2021gender, fulvio2021imbalance}. Here we sought to proactively consider choosing references that reflect the diversity of the field in thought, form of contribution, gender, race, ethnicity, and other factors. First, we obtained the predicted gender of the first and last author of each reference by using databases that store the probability of a first name being carried by a woman \cite{Dworkin2020.01.03.894378,zhou_dale_2020_3672110}. By this measure (and excluding self-citations to the first and last authors of our current paper), our references contain 23.87\% woman(first)/woman(last), 12.93\% man/woman, 11.81\% woman/man, and 51.39\% man/man. This method is limited in that a) names, pronouns, and social media profiles used to construct the databases may not, in every case, be indicative of gender identity and b) it cannot account for intersex, non-binary, or transgender people. Second, we obtained predicted racial/ethnic category of the first and last author of each reference by databases that store the probability of a first and last name being carried by an author of color \cite{ambekar2009name, sood2018predicting}. By this measure (and excluding self-citations), our references contain 14.29\% author of color (first)/author of color(last), 16.33\% white author/author of color, 15.10\% author of color/white author, and 54.28\% white author/white author. This method is limited in that a) names and Florida Voter Data to make the predictions may not be indicative of racial/ethnic identity, and b) it cannot account for Indigenous and mixed-race authors, or those who may face differential biases due to the ambiguous racialization or ethnicization of their names. We look forward to future work that could help us to better understand how to support equitable practices in science.

\begin{acknowledgments}
The authors gratefully acknowledge helpful discussions with Drs. Jordan D. Dworkin, Julia K. Brynildsen, and David M. Lydon-Staley.
\end{acknowledgments}

\bibliography{main}

\end{document}